\documentclass{iau}
\usepackage{graphicx,natbib}
 
\newcommand{\apj}{ApJ}           
\newcommand{\mnras}{MNRAS}       

\newcommand{\aj}{AJ}

\newcommand{\apjs}{ApJS}           

\title{Reconstructing the mass assembly history with kinematics and nuclear light profiles}

\author[Krajnovi\'c]{Davor Krajnovi\'c$^1$}

\affiliation{$^1$Leibniz-Institut f\"ur Astrophysik Potsdam (AIP), An der Sternwarte 16, D-14482 Potsdam, Germany. email: {\tt dkrajnovic@aip.de}}

\pubyear{2014}
\volume{311}
\jname{Galaxy Masses as Constraints of Formation Models}
\editors{M. Cappellari \& S. Courteau, eds.}

\begin{document}

\maketitle

\begin{abstract}
In this contribution I show that by combining imaging and integral-field spectroscopy it is possible to unravel the internal structure of galaxies. In particular, I will present the photometric and kinematic evidence for discs linking them with stellar angular momentum content of early-type galaxies. Furthermore, I show that the existence of both fast rotators with core and slow rotators with core-less nuclear light profiles challenges the standard formation scenarios for fast and slow rotators and suggests new pathways of mass assembly.  

\keywords{galaxies: elliptical and lenticular, cD - galaxies: evolution - galaxies: formation}
\end{abstract}

\firstsection
\section{Introduction}

In the seminal paper \citep{1983ApJ...266...41D}, Roger Davies and collaborators showed that early-type galaxies (ETGs) have divers kinematics, ranging from fainter and rotational supported to brighter and dispersion dominated systems. This picture is effectively still with us \citep[e.g.][]{2007MNRAS.379..418C,2011MNRAS.414..888E}, although the focus has somewhat changed. In particular, the usage of integral field spectrographs and the analysis of kinematic maps enable a more detailed analysis of the internal structure of galaxies.\looseness=-2 

The ATLAS$^{\rm 3D}$ Project \citep{2011MNRAS.413..813C} is a survey of morphologically selected ETGs observable with the SAURON integral-field spectrograph \citep{2001MNRAS.326...23B} mounted on the William Herschel Telescope. The galaxies selected were brighter than -21.5 magnitude in the 2MASS K$_{s}$ band and  within 42 Mpc in order to properly exploit SAURON's short wavelength range. Out of about 900 such galaxies, 260 ETGs were selected, based purely on their morphology. The sample galaxies (or their subsets) were observed with SAURON, IRAM 30m single-dish telescope, Carma Array, Westerbork Radio Synthesis Array, the Isaac Newton Telescope (INT) and the MegaCam at the Canada-France-Hawaii Telescope. In this contribution, I will review the analysis of the SAURON kinematics maps and of ground and space based imaging, looking for evidence that can be used to constrain the assembly processes of ATLAS$^{\rm 3D}$ galaxies. \looseness=-2

\section{Regularity of kinematics and the shape of early-type galaxies}

ETGs can be divided into fast and slow rotators \citep{2007MNRAS.379..401E}, a division which to the first order follows the brightness (or mass) \citep{2011MNRAS.414..888E}. The distinction between fast and slow rotators is empirically defined in the space of the specific angular momentum calculated within one effective radius, $\lambda_R$, and the observed ellipticity (also estimated within one effective radius), such that  $\lambda_R = 0.31\times\sqrt\epsilon$. This empirical division follows another empirical finding that the velocity maps of ETGs either exhibit regular or non-regular velocity features \citep{2011MNRAS.414.2923K}. The regularity of a velocity map is measured using {\tt kinemetry} \citep{2006MNRAS.366..787K}, a generalisation of the method of isophotometry \citep[e.g.][]{1987MNRAS.226..747J} to higher order moments of the line-of-sight velocity distribution (LOSVD). For the velocity maps, the best fitting ellipse is parameterised by $V=V_{rot} \cos\theta$, where $V_{rot}$ is the amplitude of rotation and $\theta$ is the eccentric anomaly. Deviations from the cosine law, parameterised with the higher order Fourier harmonics, can be quantitatively measured \citep{2008MNRAS.390...93K} and used to separate regular velocity maps, such as the `spider' diagrams of the discs galaxies, from non-regular, complex looking velocity maps, often containing kinematically distinct components, strong velocity twists or even no mean rotation. \citet{2011MNRAS.414.2923K} showed that more than 80\% of ATLAS$^{\rm 3D}$ galaxies have regular velocity maps, which, using the criteria for distinction between fast and slow rotators, translates into 86\% of fast rotators.\looseness=-2 

The velocity maps can provide one further evidence of the remarkable regular structure found in the majority of ETGs. Following \citet{1991ApJ...383..112F}, we can define the {\it kinematic misalignment angle}, $\Psi$, as the angle between the photometric major axis and the orientation of the velocity map. The kinematic misalignment angle is a powerful tool to constrain the shape of galaxies, as when $\Psi$ is different from zero, the galaxy can not have an axisymmetric shape. In \citet{2011MNRAS.414.2923K}, we measured the photometric position angle at large radii ($\sim2.5$ effective radii) and the kinematic position angle within the SAURON velocity maps (typically within one effective radius) and found that galaxies with regular velocity maps (fast rotators) are typically nearly axisymmetric systems (their $\Psi \lesssim 5^{\rm o}$). On the other hand, galaxies with non-regular  velocity maps (slow rotators) show strong kinematic misalignments (up to $90^{\rm o}$), suggesting many of them are triaxial systems. \looseness=-2

\section{Photometric and kinematics evidence for discs}

Are the regular velocity maps and the axisymmetric structure of the vast majority of ETGs a consequence of hidden discs? What fraction of the ETG population actually can be considered disc dominated galaxies? The first step in answering these questions can be achieved by analysis of images. In \citet{2013MNRAS.432.1768K}, we decomposed the surface brightness profiles, extracted from the SDSS \citep{2009ApJS..182..543A} and INT \citep{2013MNRAS.432.1894S} images, by fitting two component models consisting of a \citet{1968adga.book.....S} fitting function representing the bulge, and an exponential function, representing a disc. From this decomposition we excluded galaxies with (obvious) bars, as they would require a three component model and bars live in discs. The results show a general correspondence between fast/slow division and the existence of discs: discs are typically found and, actually, dominate fast rotators. Light profiles of some slow rotators, however, suggest embedded and small discs could be found in this population as well. A remarkable result is that the distribution of the disc-to-total light ratios (D/T) in fast rotators covers essentially the full range, from small disc contributions to fast rotators which are essentially lacking a bulge-like component\footnote{A number of  light profiles of fast rotators are best fitted with only a single S\'ersic function, but the S\'ersic index n is typical small ($<2-3$).}.\looseness=-2

\begin{figure}
\centering
\includegraphics[width=0.38\textwidth]{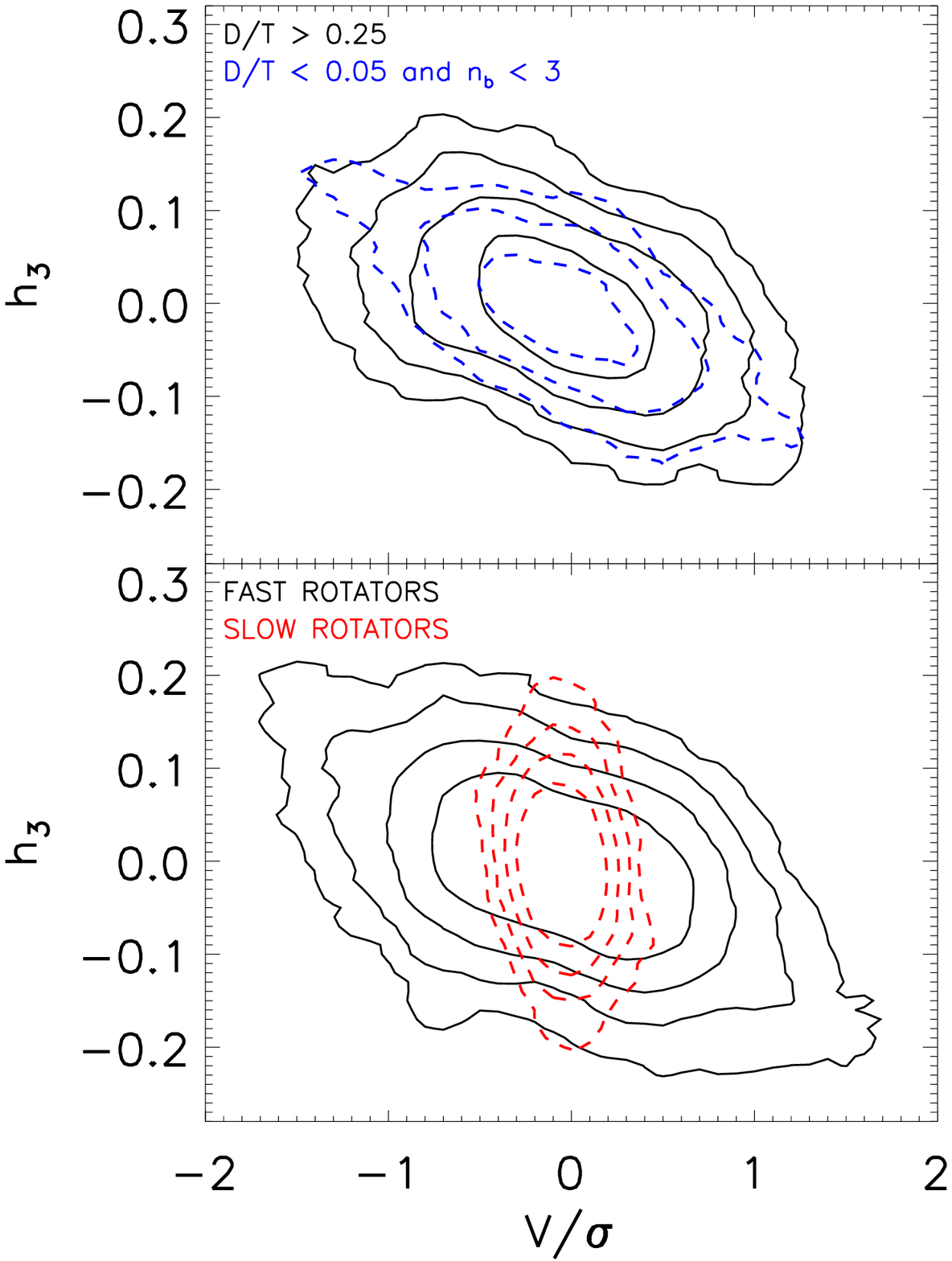} 
\includegraphics[width=0.52\textwidth]{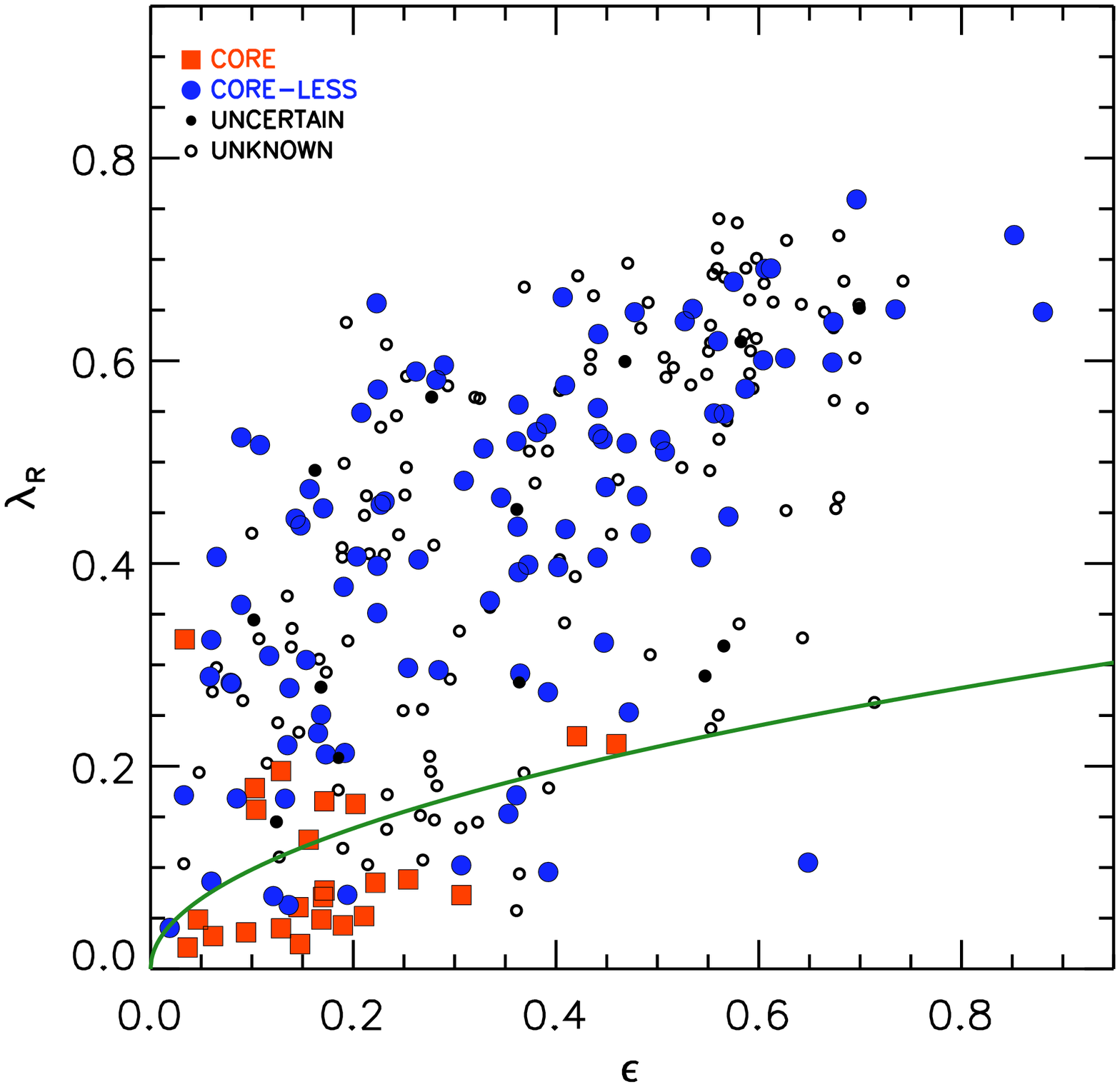} 
\caption{{\it Left:} Local $h_3 - V/\sigma$ relation for all ATLAS$^{\rm 3D}$ galaxy spectra with $\sigma>120$ km/s and an error on $h_3 < 0.05$. Top panel shows the contours for galaxies with significant disc contribution or galaxies with single S\'ersic profiles of a low S\'ersic index (dashed). Bottom panel shows the distribution for fast and slow rotators (dashed). {\it Right:} Angular momentum vs ellipticity for the ATLAS$^{\rm 3D}$ sample. The green solid line separates fast (above) and slow (below) rotators. Various symbols show the properties of nuclear light profiles, as stated on the legend. }
\label{f:fig}
\end{figure}

Due to the deprojection effects, finding discs in the photometry is essentially degenerate \citep{1990ApJ...362...52R, 1996MNRAS.279..993G}. It should be, however, more likely to determine the internal structure from kinematics. To demonstrate this, I will here make use of the third Gauss-Hermite coefficient, $h_3$, in the parametric expansion of the LOSVD \citep[e.g.][]{1993ApJ...407..525V}. This parameters measure the asymmetric deviations of the LOSVD from a Gaussian and it was shown to tightly anti-correlate with the mean velocity of the same LOSVD \citep[e.g.][]{1994MNRAS.269..785B}. This anti-correlation is a direct consequence of the existence of fast moving stars, as one would expect in an embedded disc. In galaxies which have a significant disc component or have a single S\'ersic profile of a low ($<3$) index, the anti-correlation produces a specific shape in the $h_3 - V/\sigma$, as shown on Fig.~\ref{f:fig} (top left). The same general shape is found in fast rotators, but a very different shape (no anti-correlation) is seen for slow rotators (Fig.~\ref{f:fig}; bottom left). The implication is simple: fast rotators are indeed disc dominated systems. Discs, whether large or small and embedded, are present in the majority of ETGs, and support a new classification scheme \citep{2011MNRAS.416.1680C}, which should substitute the classical Hubble diagram. \looseness=-2

\section{Angular momentum and nuclear light profies}

Nuclear surface brightness profiles of massive galaxies are known to show a bend downwards or a deficit of light with respect to a S\'ersic fit at large radii \citep[e.g.][]{1995AJ....110.2622L,2003AJ....125.2951G}. Galaxies with such profiles are often said to have {\it cores}, while galaxies whose profiles stay consistent with the outer S\'ersic fit or even show a nuclear excess of light, we will call here {\it core-less}. Core galaxies are typically more massive \citep{1997AJ....114.1771F}, as much as are slow rotators and a natural question is whether there is a clear correspondence between slow rotators and cores. In \citet{2013MNRAS.433.2812K}, we investigated this issue by matching the observations in the HST archives with ATLAS$^{\rm3D}$ galaxies. To a first approximation, fast rotators are core-less galaxies, while slow rotators are cores, but there are some exceptions (Fig~\ref{f:fig}, right). Notably there is a population of fast rotators with cores. They are kinematically consistent with being fast rotators with large discs (e.g. regular kinematics, $\Psi \approx 0^{\rm o}$, signifiant $h_3 - V/\sigma$ anti-correlation), but have $\lambda_R < 0.25$ and they are not far from the line separating fast and slow rotators. There are also some core-less slow rotators, while for a number of slow rotators their nuclear profiles are not know. Even if one removes some special cases among slow rotators\footnote{For example, recent mergers, or galaxies made up of two counter-rotating discs.}, it is obvious there is no 1:1 correspondence between cores and slow rotators and core-less and fast rotators.\looseness=-2

\section{Discussion}

What are core-less slow rotators and how are they made? How is it possible to make, or preserve, a core in a disk dominated fast rotator? These types of galaxies clearly require special formation paths. If the cores are created by the merging black holes of similar masses (brought together by a major merger event), while the core-less (light excess) profiles are made in nuclear starburst events, a fine tuning between the duration of the starburst and the binary evolution is necessary for the creation of both core fast rotators or core-less slow rotators. Creating the cores in fast rotators will require a longer evolution of the BH binary than the duration of the starburst, while the cores in the slow rotators need to be refilled by new material after the black holes are merged. This means that slow rotators can be produced in dissipative, while fast rotators in dissipation-less merges. There are indication that this is indeed the case \citep{2011MNRAS.416.1654B, 2014MNRAS.444.3357N}, but simulating the observed galaxies still requires some work.\looseness=-2

\section*{Acknowledgements}

\noindent
I thank the organisers for a very interesting meeting and the members of ATLAS$^{\rm 3D}$ for making this project a success. 


\end{document}